\documentclass[
showpacs,
 amsmath,amssymb,
prl,
floatfix,
twocolumn
]{revtex4-1}

\usepackage{graphicx}
\usepackage{bm}

\begin{document}

\title{
Emergence of coupling-induced oscillations and broken symmetries
in heterogeneously driven
nonlinear reaction networks
}

\author{Varsha Sreenivasan}
\author{Shakti N. Menon}%
\author{Sitabhra Sinha}
\affiliation{%
The Institute of Mathematical Sciences, CIT Campus, Taramani, Chennai 600113, India
}

\date{\today}

\begin{abstract}
Many natural systems including the brain comprise coupled
non-uniformly stimulated elements. In this paper we show that
heterogeneously driven networks of excitatory-inhibitory units exhibit
striking collective phenomena, including spontaneous oscillations upon
coupling. On varying the coupling strength a novel transition is seen
wherein the pattern symmetries of stimulated and unstimulated groups
undergo mutual exchange. The system exhibits coexisting chaotic and
non-chaotic attractors - an intriguing result in view of earlier
reports of varying degrees of chaoticity in the brain.
\end{abstract}

\maketitle

Complex patterns are observed to spontaneously emerge across a wide
range of spatial and temporal scales in nature~\cite{Ball1999}.
Uncovering the fundamental mechanisms driving such pattern formation will
contribute significantly towards understanding self-organization in
non-equilibrium systems~\cite{Cross1993}. Perhaps the most influential
paradigm for explaining this,
particularly in the biological context,
is that of reaction-diffusion
systems~\cite{Murray1988,Koch1994,Tlidi94,Kondo2010,Bansagi2011,Kapral2012}. 
The basic principle,
involving the interplay between self-activation and lateral inhibition
mediated by diffusion, has been suggested to underlie the emergence of
patterns in a broad range of
contexts~\cite{DeWit1996,Vanag2004,Nakao10,SinghPRE}.
However, not all phenomena involving
activator-inhibitor interactions arise through diffusive coupling, one
of the best-known examples being populations in neighboring ecological
habitats coupled through intra-specific
competition~\cite{Gilpin1973,Wu1985}. Indeed, reaction-diffusion
processes can be seen as a subset of the wider class of systems
involving nonlinear interactions between spatially distributed
elements. Novel dynamical transitions may accompany the breaking of 
any of the symmetries that are intrinsic to these systems.
Thus, uncovering the diverse range of collective
phenomena associated with non-diffusively coupled systems of
activator-inhibitor units can contribute towards understanding how
patterns can arise in a more general setting.

Neurobiological phenomena arising in synaptically coupled neuronal
populations provide some of the most varied and complex instances of
nonlinear interactions resulting in spatiotemporal
patterns~\cite{Friston2000}. Indeed, such coordinated collective
activity is seen across several spatial scales in the brain: from the
network of cortical areas where brain regions comprising $10^3$-$10^6$
neurons~\cite{Palm1993,Johansson2007} interact with each other through fiber
tracts~\cite{Modha2010}, to the olfactory bulb, where around $10^3$
glomerular clusters coordinate the information received from sensory
neurons at the nasal epithelium~\cite{Mombaerts1996}.  Each
glomerulus, which comprises circuits of excitatory and inhibitory
neurons, interact with other glomeruli through
interneurons~\cite{Lowe2013} giving rise to lateral
inhibition~\cite{Imai2014} and excitation~\cite{Christie2006}. As
these interneurons are subject to turnover~\cite{Lledo2008}, the
strength of inter-glomerular coupling can vary.  Thus, this complex
biological process can be potentially understood in terms of the
collective dynamics of a network of excitatory-inhibitory units coupled
nonlinearly with tunable strength~\cite{Migloire2015}. As each
glomerulus is activated by a specific odorant receptor
type~\cite{Mombaerts2006} and different smells evoke responses in
different combinations of glomerular clusters~\cite{Malnic1999}, it is
important in this context to understand the implications of
non-uniform stimulation on the global behavior of such a system. 

\begin{figure}[!htb]
\centering
\includegraphics[width=\linewidth]{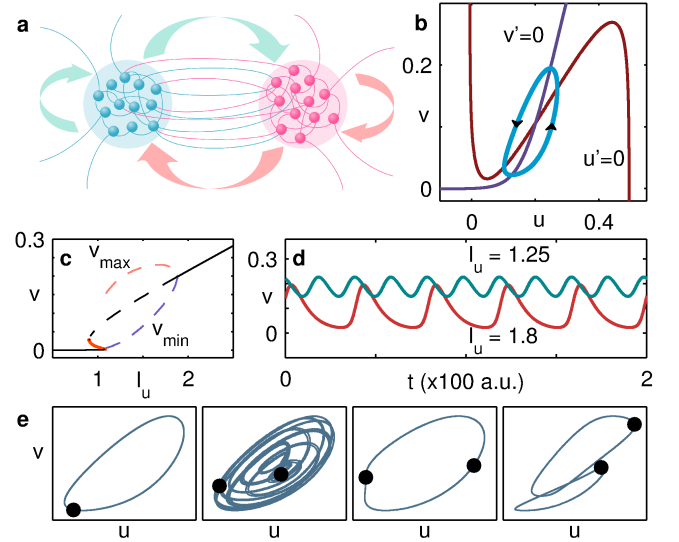}
\caption{(a) Schematic representation of a dynamical element (node) of the
network, showing the interactions between subpopulations of
excitatory and inhibitory units (neurons). (b)
Nullclines governing the dynamics of a node receiving stimulus
$I_{u}=1.25$ along with the resulting limit cycle attractor.
(c) Bifurcation diagram for the inhibitory component $v$ of an isolated
node shown as a function of the stimulus $I_u$. The broken lines indicate the
unstable branch (black), as well as the peaks ($v_{\rm max}$, pink)
and troughs ($v_{\rm min}$, violet) in the oscillatory regime. The solid black 
and thick red curves indicate the stable and saddle branches, respectively.
(d) The $v$ time-series of an isolated node
receiving stimuli $I_{u}=1.25$ (green) and $I_{u}=1.8$ (red). 
(e) Phase-plane portraits for a pair of identically stimulated 
coupled nodes (i.e., $N_{\rm stim}=N=2$) for different $w$ and
$I_{u}$ showing [L-R] exact synchronization (ES),
quasiperiodicity (QP), anti-phase synchronization (APS) and inhomogeneous in-phase synchronization (IIS). The positions of the oscillators are denoted by black filled
circles.
}
\label{fig:fig1}
\end{figure}

In this paper, we investigate the collective dynamics resulting from 
non-uniformly driven networks of identical nodes, each comprising
excitatory and inhibitory subpopulations. The heterogeneous stimulation
is implemented through external inputs being applied only to a subset
of the nodes. This conceptual framework 
applies beyond the context of 
neurobiology to phenomena as diverse as ecological interactions between
prey and predator populations~\cite{May2007} and interdependencies between
institutions in economic systems~\cite{May08,Schweitzer09}.
To describe the dynamics of the individual nodes, we consider a model
for interacting excitatory and inhibitory subpopulations
[Fig.~\ref{fig:fig1}~(a)] proposed by
Wilson and Cowan~\cite{Wilson72} obtained by temporal coarse-graining
of neuronal population dynamics.
Individual units of this type are capable of exhibiting oscillatory
activity for a range of external stimuli strength. In order to focus
on the effect of heterogeneous stimulation, we consider the simplest
connection topology, viz., coupling within and between the
subpopulations of all nodes~\cite{Singh2016}. One of the novel features
we observe is the occurrence of coupling-induced oscillations. Thus,
stimuli that generate only steady-state behavior in isolated
nodes can drive a network into oscillatory behavior, suggesting that
lateral connections between glomerular clusters can 
allow the network to recognize weak signals incapable of initiating
activity in an isolated cluster.
Increasing the strength of coupling between the nodes results in a
variety of transitions in the collective dynamics of the network, the
most striking of which involves an exchange of symmetry between
the synchronization states of the groups of stimulated and 
unstimulated nodes. In addition, we observe that the network can converge
to qualitatively distinct attractors for identical system parameters,
exhibiting chaotic or non-chaotic activity depending only on the
initial state. This result is intriguing in light of
observations reporting chaotic activity in the olfactory bulb, which
has been suggested to have a functional role in sensory information
processing~\cite{Freeman91}.

We consider a network of $N$ nodes, with each node $i$ comprising
excitatory ($u$) and inhibitory ($v$) components,
whose activity evolves as~\cite{Wilson72}:
\begin{gather}
\tau_u\frac{du_i}{dt}=-u_i + (\kappa_u - r_uu_i)\, \mathcal{S}_u({u_i}^{in}),\\
\tau_v\frac{dv_i}{dt}=-v_i + (\kappa_u - r_vv_i)\, \mathcal{S}_v({v_i}^{in}),
\end{gather}
where $\tau_{\mu}$ are time constants ($\mu=u, v$), ${u_i}^{in}$ and
${v_i}^{in}$ are the inputs received by the respective components,
$\mathcal{S}_{\mu} (x)=\kappa_{\mu} - 1 + [1 + \exp\left\{-a_{\mu}(x -
\theta_{\mu})\right\}]^{-1}$
is a sigmoidal response function
with maximum value $\kappa_{\mu}=1 - [1 + \exp\left\{a_{\mu}
{\theta}_{\mu}\right\}]^{-1}\,,$ and $r_{\mu}$, $a_{\mu}$ and
$\theta_{\mu}$ are system parameters.
As mentioned earlier, the network is globally coupled (i.e., every
node has $k=N-1$ links) with each link having the same weight $w/k$.
This normalization allows our results to be system-size
independent. The
total inputs to each component of node
$i$ are
${u_i^{in}}=c_{uu}u_i - c_{uv}v_i + \frac{w}{k}\,\sum_j (u_j - v_j) + I_{u_i}$ 
and
${v_i^{in}}=c_{vu}u_i - c_{vv}v_i + \frac{w}{k}\,\sum_j (u_j - v_j) + I_{v_i}$, 
where $j = 1, \ldots, N$ ($j \neq i$).
To implement heterogeneous stimulation different
external inputs ($I_{u_i}$, $I_{v_i}$) can be
applied to different nodes. For the results shown here we have used the 
following set of parameter values:
$c_{uu}$=16, $c_{uv}$=15,
$c_{vu}$=12, $c_{vv}$=3, $a_u$=1.3, $a_v$=2, $\theta_u$=4,
$\theta_v$=3.7, $r_{u,v}$=1, $\tau_{u,v}$=8 and $I_v = 0$. 
We have verified that our results are robust with respect
to changes in these parameter values.

\begin{figure}[!tbp]
\centering
\includegraphics[width=\linewidth]{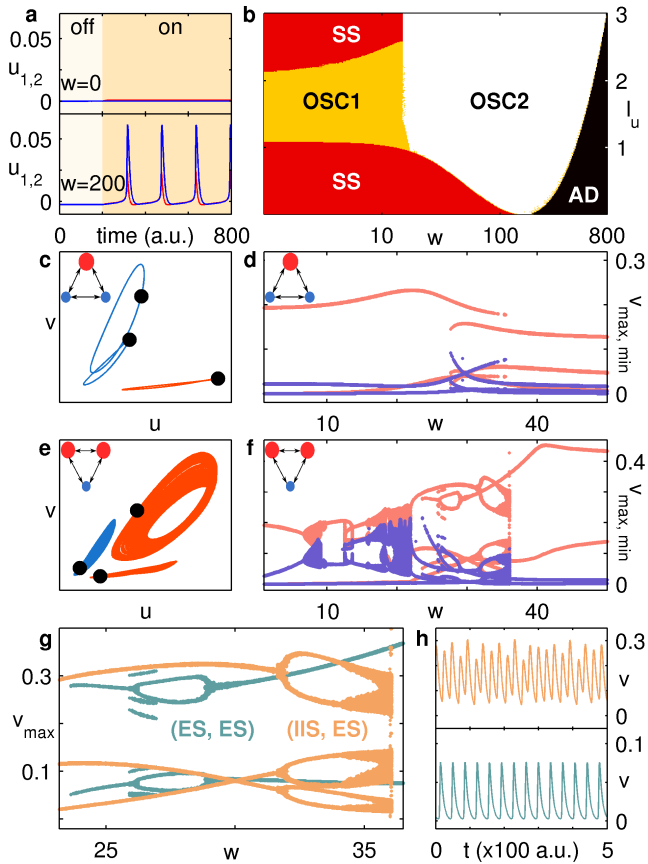}
\caption{(a-b) Coupling-induced oscillations in a pair of nodes, where
only one receives a stimulus $I_{u}$. (a) By switching ``on'' a stimulus 
that is too weak
($=0.1$) to generate activity in the uncoupled nodes (top),
oscillations can be observed by strongly coupling the
stimulated (red) and unstimulated (blue) nodes (bottom). 
(b) The range of $I_{u}$ for which oscillations emerge
increases at higher $w$. The amplitude of the oscillations is
larger for the stimulated node at low $w$ (OSC1) and for the unstimulated
node at high $w$ (OSC2). The other dynamical regimes observed
correspond to a non-zero steady state (SS) and a quiescent state
characterizing amplitude death (AD).
The regimes are determined via order
parameters and identified as the pattern obtained from the majority
($>50\%$) of random initial states (see Supplementary
Information).
(c-h) Symmetry breaking in a system of
$N=3$ globally coupled nodes. The stimulated (large,red) and
unstimulated (small, blue) nodes are indicated schematically
in top-left corner of (c-f).
Phase space projections of the trajectories (colored as per the
schematic) are shown for (c) $N_{\rm stim}=1$, displaying (ES, IIS)
state (note the broken symmetry in unstimulated nodes) and (e) $N_{\rm
stim}=2$, exhibiting chaos.
In the corresponding bifurcation diagrams for (d) $N_{\rm stim}=1$ and
(f) $N_{\rm stim}=2$,
the peaks $(v_{\rm max}$, pink) and troughs ($v_{\rm min}$, violet) of 
the inhibitory component are shown as a function of $w$.
(g) Magnified view of (f) showing the coexistence of qualitatively
distinct dynamical attractors corresponding to (IIS, ES) [orange] and
(ES, ES) [green]. The system can exhibit either chaotic or non-chaotic
behavior depending on its initial state, as illustrated in the top and
bottom panels of (h) for $w=35.6$.
}
\label{fig:fig2}
\end{figure}

On receiving a stimulus $I_{u}$ of sufficient magnitude, a single node
is capable of exhibiting limit-cycle oscillations around an unstable
fixed point~\cite{Wilson72} [Fig.~\ref{fig:fig1}~(b)]. This limit
cycle emerges via the collision of stable and saddle branches
[Fig.~\ref{fig:fig1}~(c)], and the amplitude of oscillation
depends on the value of $I_{u}$ [Fig.~\ref{fig:fig1}~(d)]. As shown
in Fig.~\ref{fig:fig1}~(e) and discussed in detail in
Ref.~\cite{Singh2016}, coupling identically stimulated nodes yields a
rich variety of synchronization patterns including exact
synchronization (ES), quasiperiodicity (QP), anti-phase
synchronization (APS) and inhomogeneous in-phase synchronization (IIS)
at different $w$ and $I_{u}$.

%
In this work, we consider heterogeneously driven networks wherein the number of
nodes receiving external stimulus $N_{stim} < N$. We henceforth denote
the synchronization state in such systems through the notation
($P_{stim}$,$P_{unstim}$), where the first and second terms in the
pair correspond to the collective pattern observed in the stimulated
and unstimulated nodes, respectively. The simplest case of non-uniform
stimulation is when a single node receiving input $I_u$ is coupled to
an unstimulated node (i.e., $N = 2$, $N_{stim} =1$) with strength $w$.
At sufficiently large $w$, oscillations can be observed even for very
low $I_u$  [Fig.~\ref{fig:fig2}~(a)]. Indeed, as shown in the phase
diagram in Fig.~\ref{fig:fig2}~(b), at high $w$ periodic activity can
be observed over a much larger range of $I_{u}$ than the one
capable of inducing oscillations in an isolated node. Furthermore,
for lower (higher) $w$ the unstimulated node has lower (higher) amplitude
oscillations than the stimulated one, denoted as OSC1 (OSC2).
This extremely
surprising {\em coupling-induced periodic activity} is seen even when
$N\gg N_{stim} = 1$ (see Supplementary Information) and suggests that a node
is capable of detecting weak, subthreshold inputs when it is coupled
to one or more unstimulated nodes. Such an increase in the sensitivity
to stimuli beyond the capability of a single element is an emergent
collective property of the network
and may be understood as an effective renormalization of the
parameters governing the nodal dynamics.
The observation of oscillations at larger values of $I_u$, where an
isolated node exhibits a stable state, may also be connected to
the appearance of periodic activity in
bistable systems, e.g., 
excitable elements subjected to a sufficiently strong stimulus, upon
appropriate coupling~\cite{In2003}. 
On increasing $w$
beyond a critical value that depends on the stimulus intensity
$I_u$, the activity of all
nodes ceases, which corresponds to a state of amplitude death (AD).

A new feature, indicative of {\em spontaneous symmetry breaking}, 
appears on minimally increasing the size of the network to $N=3$
keeping $N_{\rm stim}=1$. This is manifested as IIS in the unstimulated nodes
[Fig.~\ref{fig:fig2}~(c)], neither of which directly
receive any external stimuli but are activated only by coupling with a common 
stimulated node. Nevertheless these identical nodes exhibit distinct
oscillation patterns over a range of coupling strengths
[Fig.~\ref{fig:fig2}~(d)]. 
On stimulating a second node (i.e., $N=3$, $N_{\rm
stim}=2$), another intriguing phenomenon emerges, viz.,
the {\em coexistence} of chaotic and non-chaotic attractors.
The existence of chaotic behavior [Fig.~\ref{fig:fig2}~(e)], 
which arises via a period doubling
route [Fig.~\ref{fig:fig2}~(f)] is confirmed by verifying the
existence of positive Lyapunov exponents~\cite{Wolf85}. 
Note that if we use a smaller $I_u$, chaos can be seen in the even simpler
stimulation configuration $N_{\rm stim}=1, N=3$ . 
The novel feature for $N_{\rm stim}=2$ is that, depending on initial
conditions, it is possible to see either of
two possible collective dynamical states corresponding to
(ES, ES) or (IIS, ES), the latter being a chaotic attractor.
As both the (IIS, ES) and (ES, ES) states have non-zero basin
sizes [Fig.~\ref{fig:fig2}~(g)] we can see either chaotic or non-chaotic
behavior for identical system parameters over a 
range of $w$ [Fig.~\ref{fig:fig2}~(h)]. 
This observation lends support to the hypothesis, based on
experimental recordings from the rabbit olfactory system, that
attractors with varying degrees of chaoticity can coexist in the
brain~\cite{Freeman2000}. 


\begin{figure}[!tbp]
\centering
\includegraphics[width=\linewidth]{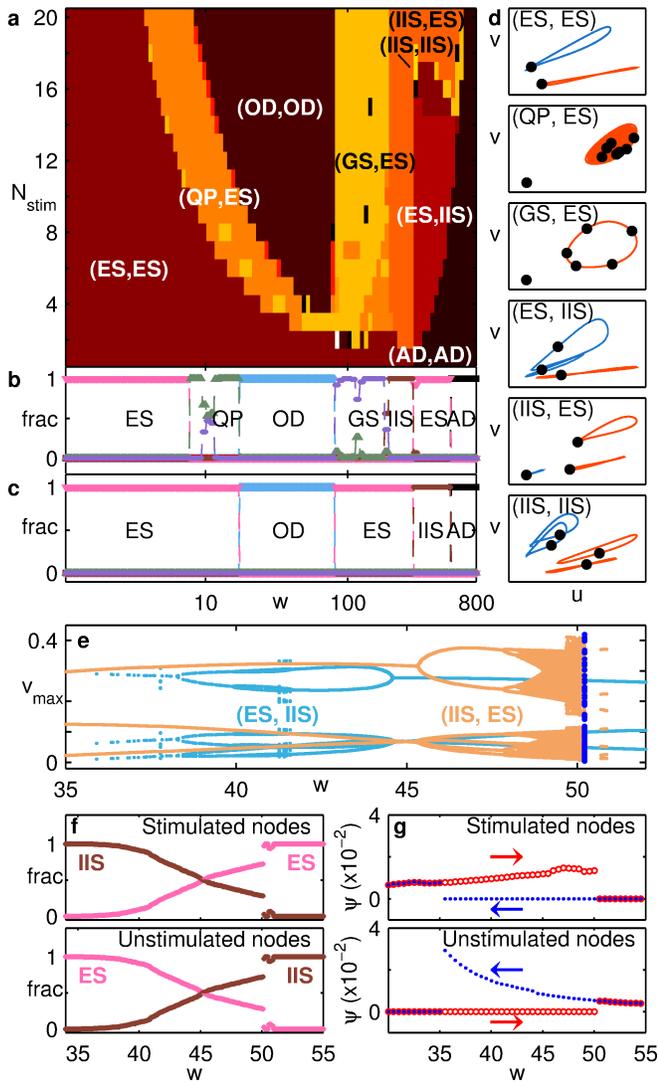}
\caption{(a-d) Collective dynamics of a system of $N=20$ globally
coupled nodes. (a) Different synchronization states obtained by varying the
number of stimulated
nodes, $N_{\rm stim}$, and coupling strength, $w$,
with (${\rm P}_{stim}$, ${\rm P}_{unstim}$) referring to patterns in
stimulated and unstimulated groups.
(b-c) Variation of the attraction basin
size
(measured as fraction of initial states reaching the attractor) with
$w$ at $N_{\rm stim}=10$ 
for the different regimes in (a), shown separately for the (b)
stimulated and (c) unstimulated groups.
%
Basin sizes have been estimated using $10^2$ initial conditions.
(d) Phase space projections of the trajectories (red: stimulated,
blue: unstimulated)
corresponding to the different synchronization states indicated in (a). 
(e-g) Symmetry exchange transition in a system of $N=4$ globally
coupled oscillators with
$N_{\rm stim}=2$. (e) Bifurcation diagram showing the peaks $(v_{\rm max})$
of the inhibitory components of all nodes as $w$ is varied. 
Distinct coexisting attractors corresponding
to (IIS, ES) and
(ES, IIS) are indicated by orange and light blue, respectively.
(f-g) Variation of the (f) fractional basin size of these attractors
and (g) the order parameter $\psi = \left<{\sigma_{i}}^2(v_i)\right>$
with $w$, shown
separately for
the (top) stimulated and (bottom) unstimulated
groups. The distinct trends seen in (g) on increasing (red circle) or
decreasing (blue dots) $w$ indicate the occurrence of hysteresis-like
behavior.
}
\label{fig:fig3}
\end{figure}

%
The existence of numerous synchronization regimes in the $w-N_{\rm
stim}$ parameter space becomes apparent as we increase the
complexity of the system by incorporating more  nodes
[Fig.~\ref{fig:fig3}~(a)].
These regimes are typically demarcated by sharp
changes in the sizes of the basin of attraction of individual patterns
indicated in Fig.~\ref{fig:fig3}~(b-c). 
Apart from the states corresponding to ES, QP, IIS and AD
described earlier, new collective dynamical patterns for the
stimulated and unstimulated groups emerge. These
include oscillator death (OD) which is a homogeneous non-zero steady
state, and gradient synchronization (GS), a generalization of APS for
systems with $N > 2$~\cite{Singh2016}. 
Fig.~\ref{fig:fig3}~(d) shows several of the possible collective
patterns, including those corresponding to symmetry breaking (IIS) in one
or both groups of stimulated and unstimulated nodes. 
A particularly surprising feature that we investigate in detail below
is the existence of a novel transition in which the (broken) symmetry of the
patterns in the stimulated and unstimulated groups undergo a mutual
exchange on varying the coupling strength. This symmetry exchange
manifests as a transition from the (IIS, ES) to the (ES, IIS) state
[Fig.~\ref{fig:fig3}~(a)].

We examine the nature of this transition in detail by considering the
simplest system in which it can be observed. To observe broken
symmetry in each group, both the stimulated and
unstimulated groups should contain at least two nodes each.
We observe that in the minimal case, i.e., for $N=4$, $N_{\rm
stim}=2$, the states (IIS, ES) and (ES, IIS) co-exist over a range of $w$
[Fig.~\ref{fig:fig3}~(e)]. The transition between them occurs through
a change in the relative basin sizes for the patterns in the two
groups [Fig.~\ref{fig:fig3}~(f)]. 
The mechanism of the (broken) symmetry exchange is further established
by examining how the order parameter $\psi=\left<{\sigma_{i}}^2(v_i)\right>$
(non-zero values of which indicate IIS in this regime) changes upon varying
$w$ in either direction in an annealed manner. In this procedure, the
system is allowed to
evolve starting from a random initial state at low (high) $w$,
following which the value of $w$ is adiabatically increased
(decreased). As seen in Fig.~\ref{fig:fig3}~(g), a hysteresis-like
behavior can be observed in the transition region in both groups of
nodes, consistent with the mechanism of shrinking basin sizes
underlying the symmetry exchange.

Our results provide a simple framework for understanding aspects
of the complex patterns of spatiotemporal activity that the brain is
observed to exhibit~\cite{Wilson2006}. 
The mesoscopic approach used here, focusing on
the dynamics of a network of neuronal clusters, can yield significant
insights into the mechanisms by which such patterns emerge.
Furthermore, the phenomena we report here occurs in a globally connected
system, which may help elucidate the important role of long-range
intracortical interactions in olfactory processing~\cite{Luo2011}.
Among the variety of dynamical
transitions observed upon varying the coupling strength, the most striking 
one involves an exchange of broken and restored symmetries
between the groups
of stimulated and unstimulated nodes.
This result suggests an experimentally testable
hypothesis, namely that
simulated and non-stimulated glomeruli may show distinct collective
dynamics at different levels of arousal which are analogous to having 
different coupling strengths between nodes.

To conclude, we have shown that non-uniformly driven networks of
identical
nodes, each comprising excitatory and inhibitory subpopulations, are
capable of exhibiting surprisingly rich collective phenomena.
We find that the system exhibits {\em coexistence of qualitatively
distinct attractors}
(chaotic and non-chaotic) for identical system parameters.
This is intriguing in light of experimental
observations of chaotic dynamics of varying complexity, particularly in
the olfactory system~\cite{Freeman91,Freeman2000,Korn2003}, reported
over several decades and which are yet to be fully understood.
Furthermore, nodes that are quiescent in isolation can
spontaneously oscillate for sufficiently strong coupling, thus
enabling
the system to be activated even by stimuli that are incapable of
generating dynamical activity in isolated nodes. These
{\em coupling-induced oscillations} may help in
understanding how excitatory-inhibitory feedback can cause spontaneous
oscillations in networks of naturally quiescent stochastic spiking
neurons~\cite{Wallace2011}, 
whose mean-field limit can in fact be described through the
Wilson-Cowan equations.
Moreover, this
can be understood as
belonging to a larger class of phenomena characterized by the emergence 
of activity in
quiescent systems upon coupling~\cite{Taylor2009,Singh2012,Xu2015}.
In summary, our results suggest that transitions between symmetry broken
and restored symmetry states in heterogeneously driven system of
coupled neural oscillators may underlie the sequence of complex
activity patterns seen in the brain.

We would like to thank Bhaskar Saha, Rajeev Singh, Sudeshna Sinha,
Deepak Dhar and K.~A.~Chandrashekar
for helpful discussions. SNM is supported by the IMSc Complex Systems
Project ($12^{\rm th}$ Plan). VS is supported by the ITRA Project. We
thank IMSc for providing access to the supercomputing cluster
``Satpura'', which is partially funded by DST.


\pagebreak
\onecolumngrid
\begin{center}
{\large \bf SUPPLEMENTARY INFORMATION}
\end{center}
\setcounter{figure}{0}
\renewcommand\thefigure{S\arabic{figure}}
\renewcommand\thetable{S\arabic{table}}

\section{Order parameters}
\begin{figure*}[!htb]
\includegraphics[scale=0.99]{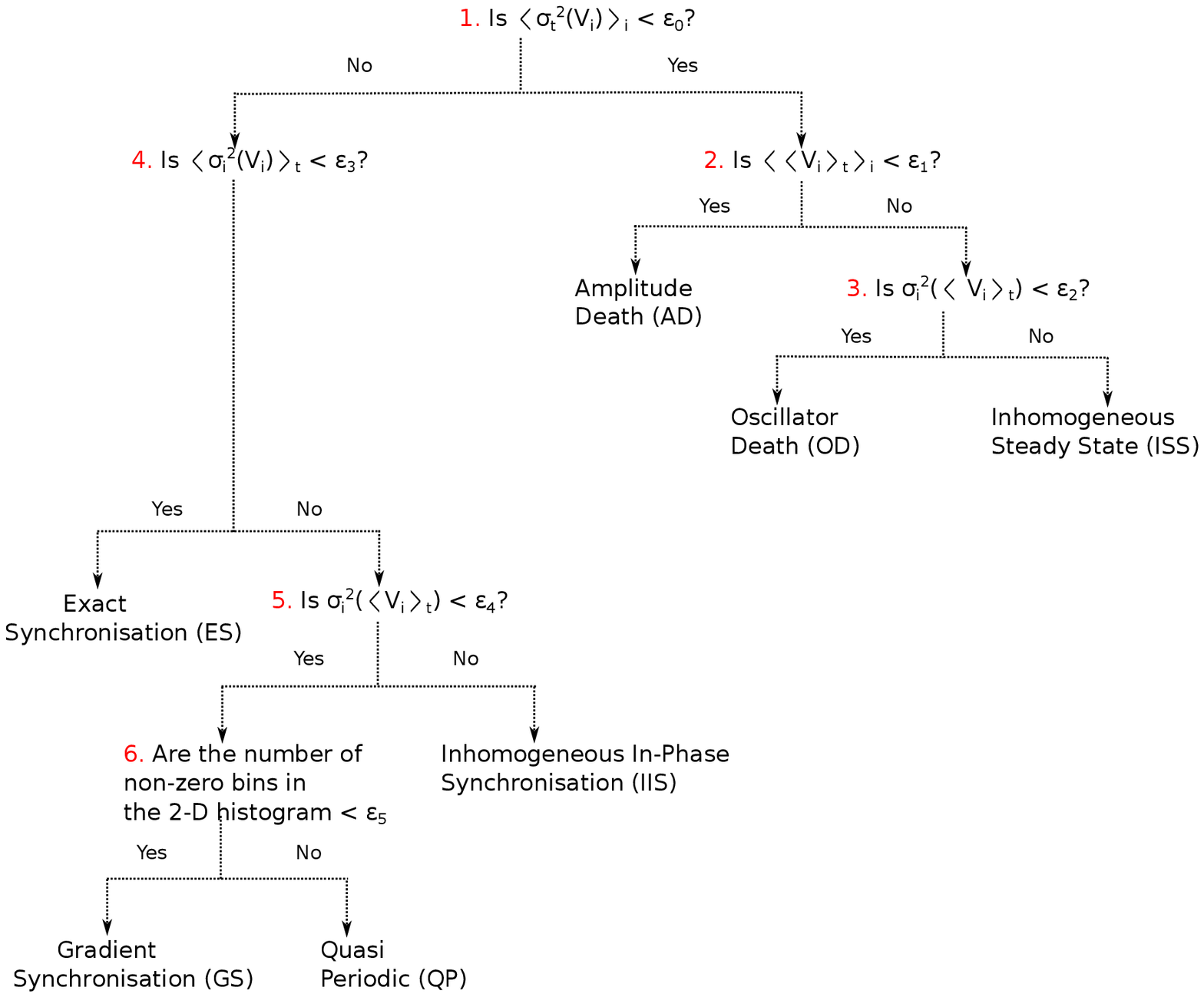}
\caption{Decision tree for the use of order parameters to distinguish
between the different collective dynamical states. Although this tree
is identical for both stimulated and unstimulated oscillators, the
threshold values chosen for identifying the corresponding patterns
are different (see Table~\ref{table:op_table}).}
\label{fig:order_parameters}
\end{figure*}

The identification of each collective dynamical state is done by
measuring a set of order parameters and then using the
decision tree in Fig.~\ref{fig:order_parameters} to classify the
synchronization state. The same procedure is applied independently for
the
stimulated and the unstimulated groups of nodes. At each numbered
decision point, threshold values (Table \ref{table:op_table}) for the
order parameters are
employed to answer the following questions:
\begin{enumerate}
\item Whether the magnitude of the inhibitory component of each
oscillator changes over time. For a non-oscillating state, this
quantity should be below a threshold $\varepsilon_0$.
\item Whether the magnitude of the inhibitory component of each
oscillator is close to zero for all oscillators. If this quantity is
below a threshold $\varepsilon_1$, it signifies amplitude death.
\item Whether the temporal mean of each oscillator is the same.
If the difference is below a threshold $\varepsilon_2$, it signifies
oscillator death; otherwise it is an inhomogeneous steady state.
\item Whether all the oscillators are in phase. If the difference
between the phases of the oscillators is less than a threshold
$\varepsilon_3$, it corresponds to exact synchronization.
\item Whether the temporal mean of each oscillator is the same.
If the difference is greater than a threshold $\varepsilon_4$, it
signifies
inhomogeneous in-phase synchronization.
In the oscillating case, once phase synchronization has been
established, this order parameter checks if all oscillators have the
same trajectory.
\item Whether the total amount of phase space covered by the
trajectory is small. For the case of GS, this quantity is less than a
threshold $\varepsilon_5$; if the quantity is greater than the
threshold, the state is identified as QP which, by definition,
completely fills a bounded region in phase space.
\end{enumerate}
\begin{table*}[!htb]
\begin{tabular}{|c|c|c|}\hline
Threshold & Value (Stimulated) & Value (Unstimulated)\\\hline
$\varepsilon_0$ & $10^{-7}$  & $10^{-15}$\\\hline
$\varepsilon_1$ & $10^{-10}$ & $10^{-10}$\\\hline
$\varepsilon_2$ & $10^{-10}$ & $10^{-15}$\\\hline
$\varepsilon_3$ & $10^{-9}$  & $10^{-12}$\\\hline
$\varepsilon_4$ & $10^{-4}$  & $10^{-5}$ \\\hline
$\varepsilon_5$ & $2 \times 10^{4}$   & $2 \times 10^{4}$  \\\hline
\end{tabular}
\caption{Thresholds for the order parameters used to distinguish the
synchronization patterns (see Fig.~\ref{fig:order_parameters}). To
account
for the fact that unstimulated nodes typically exhibit much smaller
oscillations, the chosen threshold values are in general different for
the
stimulated and the unstimulated groups.}
\label{table:op_table}
\end{table*}
\clearpage

\section{Robustness of results}
\begin{figure*}[!htb]
\includegraphics[width=\linewidth]{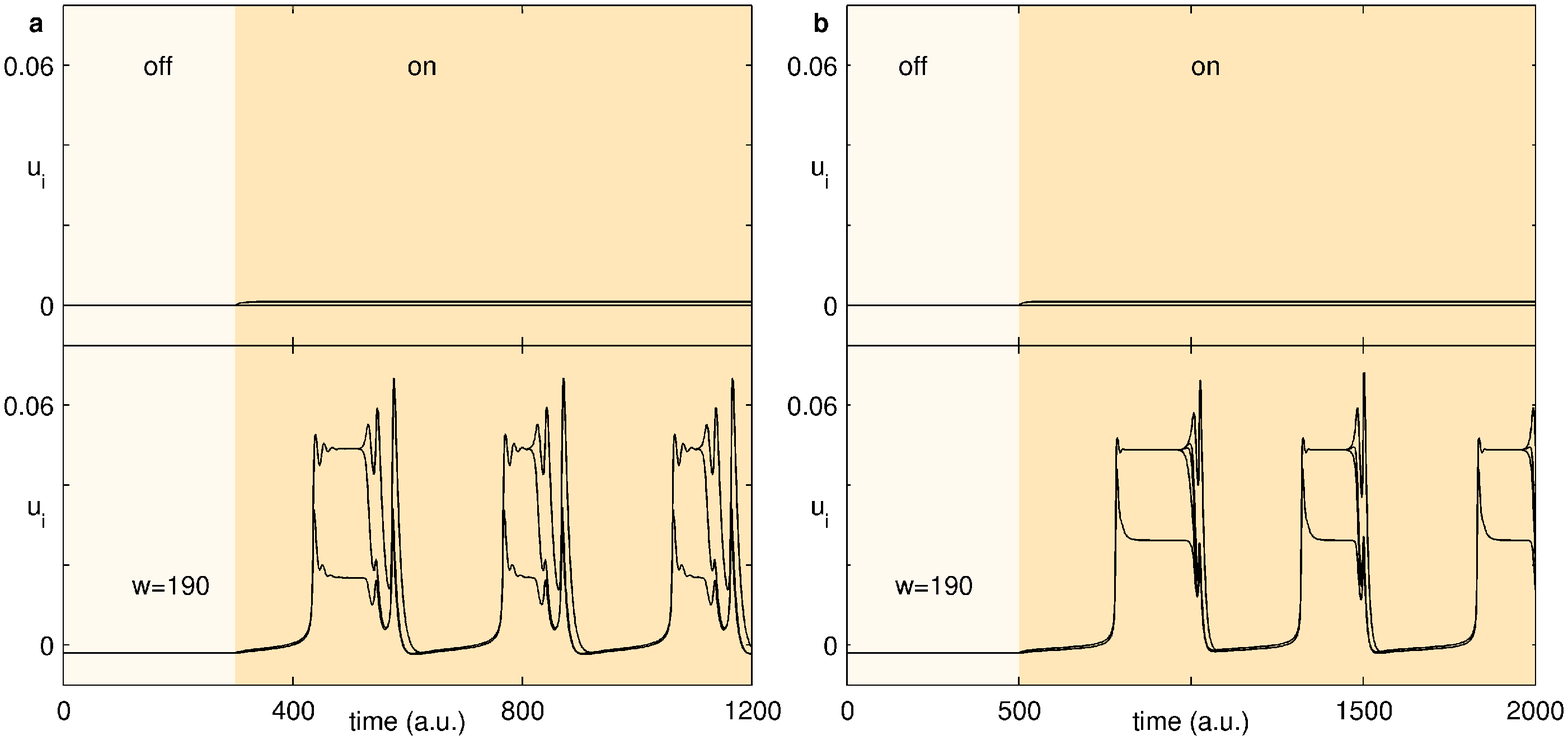}
\caption{Coupling-induced oscillations in a system of
(a) $N=3$ and
(b) $N=5$ nodes, where
only one node receives a stimulus $I_u=0.1$ that is too weak to
generate activity in the uncoupled nodes (top). Oscillations can be
observed
by strongly coupling ($w=190$) the nodes (bottom).
}
\label{fig:SI_fig_onoff}
\end{figure*}

In the following, we demonstrate that the results presented in the
main text are robust with respect to small changes in the system
parameters or system size. First, we consider the case of
coupling-induced oscillations, shown in Fig.~2~(a) of the main text.
As
seen in Fig.~\ref{fig:SI_fig_onoff}, steady state behavior is
exhibited by a globally coupled system of $N$ ($=3, 5$) nodes, with
only $N_{\rm stim}=1$ of them receiving a weak input stimulus $I_{u}$.
Although the stimulation
is too weak to initiate activity in an isolated node,
upon coupling these nodes
with sufficient strength $w$, we find that oscillations emerge, as in
the case $N=2$ considered in the main text.\\

\begin{figure*}[tbp]
\includegraphics[width=\linewidth]{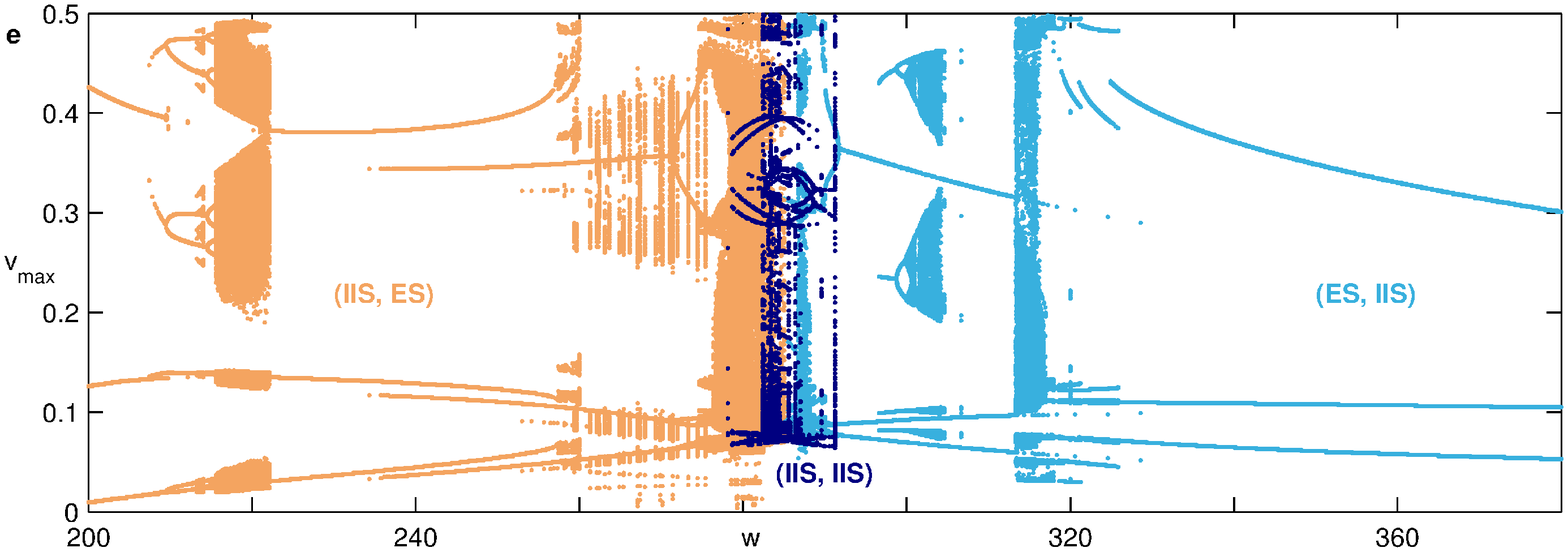}
\caption{Bifurcation diagram for a network with $N=20$, $N_{\rm
stim}=10$,
with the peaks $(v_{\rm max}$) of the inhibitory component, $v$,
indicated over a range of $w$. Colors are used to represent different
collective dynamical states, viz., orange for (IIS, ES), light blue
for (ES, IIS) and dark blue for (IIS, IIS). On increasing $w$, a
transition occurs from (IIS, ES) to (ES, IIS), corresponding to an
exchange
of (broken) symmetries between the stimulated and unstimulated groups.
}
\label{fig:SI_fig_BIF}
\end{figure*}

Next, we consider the symmetry switching transition discussed in
the main text. In Fig.~3~(e) of the main text, we see that for
the case $N=4$, $N_{\rm stim}=2$ there exists a range of values within
which the distinct collective dynamical states (ES, IIS) and (IIS,
ES) coexist. As seen in Fig.~3~(f), the likelihood of obtaining the
pattern pair (IIS, ES) is higher at lower values of coupling strength
$w$, and only (ES, IIS) is obtained above a critical value of $w$.
These results hold even for much larger system sizes, as shown in
Fig.~\ref{fig:SI_fig_BIF}, which displays a bifurcation diagram for
the case $N=20$, $N_{\rm stim}=10$. As we can see, the pattern pair
(IIS, ES) is seen for lower $w$, and (ES, IIS) is seen for higher
$w$.\\

\begin{figure*}[tbp]
\includegraphics[width=\linewidth]{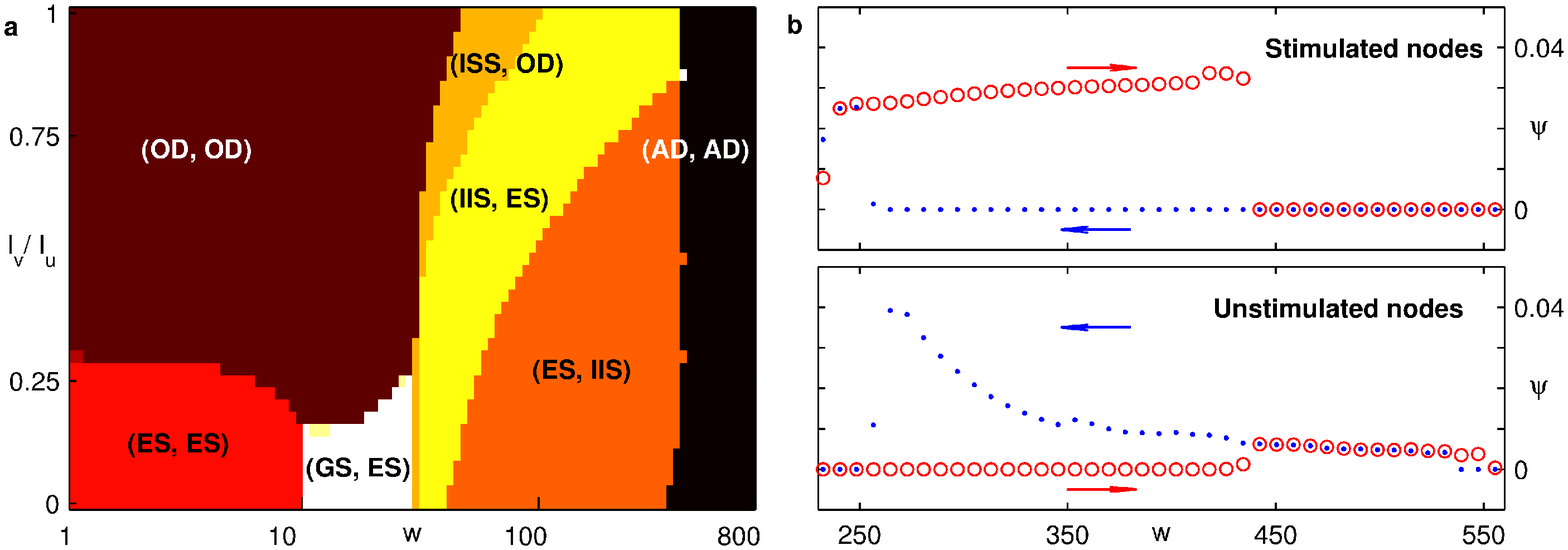}
\caption{Collective dynamics of the system on changing model
parameters, viz., halving the internal coupling strengths
$c_{\mu,\nu}$, show results qualitatively similar to those described
in the main text.
(a) Collective dynamics of a system of $N = 4$ globally coupled nodes,
with $N_{\rm stim} = 2$,
showing the different synchronization states obtained by varying the
coupling strength $w$ and the ratio of the input stimuli received by
the inhibitory and excitatory components $I_{v}/I_{u}$ (fixing $I_{u}
= 1.25$).
(b) Variation of the order parameter
$\psi=\left<{\sigma_{i}}^2(v_i)\right>$
with $w$ for a network with $N = 40$, $N_{\rm stim} = 20$ ($I_u =
1.25$, $I_v = 0$).
The behavior of the (top) stimulated and (bottom) unstimulated groups
are shown separately. The distinct
trends seen on increasing (red circle) or decreasing (blue
dots) $w$ indicate the occurrence of hysteresis-like behavior.
}
\label{fig:SI_fig4}
\end{figure*}

Finally, we show that our results are robust with respect to changes
in both the internal coupling parameters as well as the
input stimulus applied to the inhibitory components ($I_{v}$). In
Fig.~\ref{fig:SI_fig4}~(a), we show a parameter space diagram
indicating
the synchronization states obtained upon changing $w$ as well as
the ratio $I_{v}/I_{u}$, for the case $N=4$, $N_{\rm stim}=2$. Note
that, for
these simulations, the values of all the internal
coupling strengths $c_{\mu,\nu}$ have been halved relative to those
used for the simulations reported in the main text.
We find that for low values of the
ratio ($\lesssim 0.25$) the patterns observed, and their phase
boundaries, are almost unchanged from the case $I_{v}=0$. At larger
values of $I_{v}$, the phase boundaries change, but we continue to
observe the symmetry switching transition from (IIS, ES) to (ES, IIS).
Furthermore, as shown in Fig.~\ref{fig:SI_fig4}~(b), we continue to
observe
a hysteresis-like behavior in the range of coupling strengths around
the symmetry-switching transition even at large system sizes ($N=40$,
$N_{\rm stim}=20$).
These observations suggest that the results reported in the main text
are robust with respect to small variations in the parameter values
and system size.

\section{Numerical simulations}

The equations corresponding to the system of $N$ globally coupled
nodes are solved numerically using the variable order method for stiff
differential equations as implemented in the \texttt{ode15s} routine
in
MATLAB\textsuperscript{\textregistered}. The time-series data used for
analysis is recorded after allowing sufficient time for any transient
effects to dissipate (typically $2 \times 10^4$ time units).

\end{document}